\documentclass[11pt,twoside]{article}
\usepackage{./asp2014}

\aspSuppressVolSlug
\resetcounters

\begin{document}

\title{SETI Searches for Evidence of Intelligent Life in the Galaxy}
\author{Steve Croft$^1$, Andrew P.~V.\ Siemion$^{1,2,3}$, James M.\ Cordes$^4$, Ian Morrison$^5$, Zsolt Paragi$^6$, Jill Tarter$^3$
\affil{$^1$Department of Astronomy, University of California, Berkeley, 501 Campbell Hall \#3411, Berkeley, CA, 94720, USA}
\affil{$^2$Department of Astrophysics/IMAPP, Radboud University, P.O. Box 9010, NL-6500 GL Nijmegen, The Netherlands}
\affil{$^3$SETI Institute, Mountain View, CA 94043, USA}
\affil{$^4$Cornell Center for Astrophysics and Planetary Science and Department of Astronomy, Cornell University, Ithaca, NY 14853, USA}
\affil{$^5$Curtin Institute of Radio Astronomy, Curtin University, Perth, Australia}
\affil{$^6$Joint Institute for VLBI ERIC (JIVE), Postbus 2, NL-7990 AA Dwingeloo, The Netherlands}
}

\paperauthor{Steve Croft}{scroft@berkeley.edu}{0000-0003-4823-129X}{UC Berkeley}{Dept. of Astronomy}{Berkeley}{CA}{94720}{USA}
\paperauthor{Andrew Siemion}{siemion@berkeley.edu}{0000-0003-2828-7720}{UC Berkeley}{Dept. of Astronomy}{Berkeley}{CA}{94720}{USA}
\paperauthor{James Cordes}{cordes@astro.cornell.edu}{0000-0002-4049-1882}{Cornell University}{Cornell Center for Astrophysics and Planetary Science and Department of Astronomy}{Ithaca}{NY}{14853}{USA}
\paperauthor{Ian Morrison}{dr.ian.s.morrison@gmail.com}{0000-0003-0833-0541}{Curtin University}{Curtin Institute of Radio Astronomy}{Perth}{WA}{}{Australia}
\paperauthor{Zsolt Paragi}{zparagi@jive.eu}{0000-0002-5195-335X}{Joint Institute for VLBI ERIC (JIVE)}{}{Dwingeloo}{}{NL-7990 AA}{The Netherlands}
\paperauthor{Jill Tarter}{tarter@seti.org}{}{SETI Institute}{}{Mountain View}{CA}{94043}{USA}

\begin{abstract}
Radio SETI experiments aim to test the hypothesis that extraterrestrial civilizations emit detectable signals from communication, propulsion, or other technologies. The unprecedented capabilities of next generation radio telescopes, including ngVLA, will allow us to probe hitherto unexplored regions of parameter space, thereby placing meaningful limits on the prevalence of technological civilizations in the Universe (or, if we are fortunate, making one of the most significant discoveries in the history of science). ngVLA provides critical capabilities in the 10 -- 100 GHz range, and will be a valuable complement to SKA in the southern hemisphere, as well as surveying the sky at frequencies underexplored by previous SETI experiments.
\end{abstract}

\section{Description of the problem}
Almost all stars are now thought to have planets, and a substantial fraction \citep{kasting:14,dressing:15} have a planet in the ``habitable zone'' where liquid water may exist on the surface. The local Universe is thus a target-rich environment to search for signs that life may have taken hold elsewhere in our Galaxy. This can be accomplished either by looking for atmospheric biosignatures (using the next generation of space-based or extremely large ground-based optical telescopes), or by looking for advanced life by searching for signatures of technology \citep{tarter:01}. The latter will be an area in which ngVLA will excel. The search for extraterrestrial intelligence (SETI) is also complementary to other proposed ``cradle of life'' studies to be undertaken by ngVLA, including studying star and planet formation, and the formation of complex molecules \citep{ngvla_memo6}.

The frequency coverage of ngVLA is very complementary to the bands to be covered by SKA \citep{siemion:13}, allowing a comprehensive SETI search to be undertaken over three decades of radio frequency. These new instruments will also provide the sensitivity required to detect analogs of Earthbound transmitters such as aircraft radar from the distance of several nearby stars, in addition to more powerful (or directional) transmitters from across the Galaxy. 

\section{Scientific importance}
``Are we alone?'' is one of the most profound human questions. Yet at the turn of the seventh decade \citep{cocconi:59,drake:61} of SETI searches, there is, as yet, no evidence that intelligent life has arisen elsewhere in our Universe. One might infer that technological life beyond Earth is rare, or even non-existent, but in reality such a small volume of parameter space has been explored to date that this conclusion would be premature. The search for extrasolar planets provides an instructive example. As late as the 1980s, searching for planets orbiting other stars was a fruitless task, but as new techniques and powerful new instruments became available, the field has exploded \citep[e.g.,][]{borucki:10}. We now understand that the Galaxy is replete with planets of a huge variety of sizes, temperatures, and compositions, including many similar in character to our own.

\section{Astronomical impact}
The detection of the first exoplanets opened an entirely new field of research. Aside from the cultural impact of a SETI detection, the dawning of the age of observational astrobiology will mark one of the most significant milestones in the history of astronomy. Even the detection of a simple beacon, indicative of the presence of another technological civilization in our Galaxy, would enable us to begin to understand the processes by which the Universe gives rise to life, and intelligence, launching a new Copernican Revolution by which we more fully understand our place in the cosmos. Were a signal to contain information that we were able to comprehend, we could potentially understand what awaits us as a species, or begin to investigate ``the archaeology of the future'' \citep{morrison:97}.
\section{Anticipated results} 
%
Radio SETI experiments aim to test the hypothesis that extraterrestrial civilizations emit detectable signals from communication, propulsion, or other technologies. SETI attempts to constrain the terms in the Drake Equation \citep{drake:61,worden:17}; in particular, given that we now know that habitable planets are common, can we attempt to constrain the fraction of such planets that develop life, intelligence, and technology, and the lifetime over which signs of such technology are detectable? We do not know, a priori, the frequency spectrum, luminosity function, duty cycle, or antenna gain and beaming fraction of ETI transmissions. To have the best chance of success, therefore, a SETI survey must attempt to explore a large region of parameter space in frequency (coverage and resolution), time (resolution, cadence, and duration), and sensitivity. While it is challenging to qualitatively compare one SETI survey to another, increases in computational power and survey speed are enabling modern SETI searches that far exceed the capabilities of their predecessors \citep{enriquez:17}.

As a fast survey instrument with high sensitivity and wide frequency coverage, ngVLA will provide powerful constraints on the space density of technologically advanced civilizations employing transmission technologies over a wide swath of the radio spectrum, if not actually directly detecting such civilizations' existence.

\section{Limitations of current astronomical instrumentation}
\begin{figure}
\centering
\includegraphics[width=\linewidth]{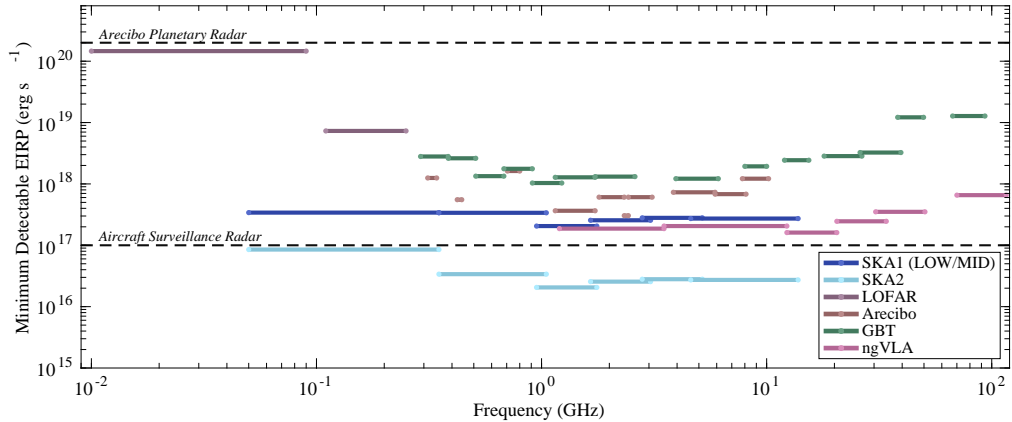}
\caption{\label{fig:sens}
 Sensitivity of the ngVLA to narrow-band transmitters at 15 pc, as compared with other
facilities actively performing SETI searches over the same band (following \citealt{siemion:13}). Here we assume a significance threshold $\sigma = 15$,
bandwidth $\Delta b = 0.5$\,Hz and integration time $t = 10$ minutes. A transmitter is detectable if its EIRP is above the curve
for a given telescope. Thus a transmitter with an EIRP of $2 \times 10^{20}$ ergs/sec (planetary radar) is detectable with all of the
telescopes shown, while a transmitter with an EIRP of $1 \times 10^{17}$ ergs/sec (airport radar) is detectable only with SKA2 at 15\,pc, but could be detected by ngVLA for stars closer than $\sim 10$\,pc.
Sensitivities for the Green Bank Telescope (GBT), Arecibo and LOFAR were taken from those facilities' observing
guides. For LOFAR we assume only core stations are used. The ngVLA sensitivity was taken from \citet{selina:18}.
}
\end{figure}

Although the power of extraterrestrial transmissions may be considerably higher than those possible with human technologies, it is nevertheless useful to use our own transmission capabilities as a guide. Our most powerful beamed transmitter, the Arecibo telescope when used as a planetary radar, has an equivalent isotropic radiated power (EIRP) of $2 \times 10^{20}$ erg/s, and would be detectable by ngVLA at distances of kiloparsecs. ETI signals comparable to our most powerful wide-angle transmitters, such as airport radars (EIRP $\sim 10^{17}$ erg/s) would be detectable from planets orbiting the nearest few thousand stars to the Sun. ngVLA will complement SKA1-LOW and SKA1-MID as the only facilities with the capability to detect ``leakage'' transmissions from omnidirectional transmitters with power close to the brightest transmitters on Earth (Figure~\ref{fig:sens}).

\section{Connection to unique ngVLA capabilities}

The unprecedented capabilities of next generation radio telescopes, including ngVLA, will allow us to probe hitherto unexplored regions of parameter space \citep{ekers:02}, thereby placing meaningful limits on the prevalence of technological civilizations in the Universe (or, if we are fortunate, making one of the most significant discoveries in the history of science).

ngVLA provides critical capabilities in the 10 -- 100 GHz range, a region of the spectrum used by many human technologies, to survey the sky at sensitivities unmatched by other facilities, and underexplored by previous SETI experiments. At frequencies where ngVLA and SKA overlap, ngVLA will primarily see the northern hemisphere and SKA the southern.

Additionally, ngVLA combines high sensitivity with very high resolution (of order 10\,mas). With such high resolution, the motion of a putative transmitter on a planet or spacecraft, orbiting its star at a distance of $\sim $AU, could be discerned astrometrically even for stars at distances of $\sim 100$\,pc. Measuring the motion of such a transmitter would be a powerful additional tool for distinguishing artificial transmitters in orbit around nearby stars.

\section{Experimental layout}  

SETI searches have two key goals:
\begin{enumerate}
\item Find signals that appear artificial (e.g. narrow band, modulated, pulsed, or otherwise obviously ``unnatural'')
\item Localize such signals at a specific right ascension and declination (to eliminate the possibility that they are radio frequency interference (RFI) from transmitters close to the telescope or from Earth-orbiting satellites)
\end{enumerate}

Searches can either be performed using baseband voltage data, or, after an FFT operation, spectrograms or ``waterfall plots'' of intensity as a function of frequency and time. Classical SETI algorithms have typically focused on the search for narrow-band signals, but increased computing power and more sophisticated algorithms (including machine learning approaches) will enable the search for increasingly complex signals, as well as better classification of confounding RFI.

We envisage an ngVLA SETI search operating commensally with most primary science users of the array, but in addition a smaller amount (tens to hundreds of hours per year) of TAC-allocated time for pointed observations in support of time-critical or other targets of interest. Any star within the instantaneous field of view of the ngVLA is a potential SETI target, and as such a commensal search with ngVLA would enable stringent limits on ETI transmissions for millions of stars, in addition to limits at higher EIRP for more distant targets such as galaxies. In addition to stars and galaxies, solar system objects have in the past also been targets for SETI searches \citep[e.g.,][]{freitas:80,omm_gbt, omm_mwa}; commensal or targeted searches of such objects with ngVLA could easily detect transmitters with EIRPs measured in milliwatts.

The proposed observations require that the ngVLA have the capability to deliver time domain voltage data from every antenna to dedicated signal processing hardware to be installed as either a user or facility instrument at ngVLA.  Similar to user instruments operating at Arecibo, Green Bank, and Parkes, such a system would run in a commensal mode, capturing or processing voltage data alongside non-SETI observations, in addition to standalone observations of SETI targets. Such equipment would consist of $\sim 100$\,PB storage and 10\,PFLOPS compute (assuming Moore's Law extrapolation from current equipment) enabling digitization of the entire ngVLA bandwidth and be capable of forming and searching both incoherent and coherent (phased) beams. 

As well as operating commensally with primary science programs, flexible sub-arraying capabilities would allow some of the antennas to undertake a dedicated SETI search simultaneously with other users of the array (albeit at reduced sensitivity). When operating in a commensal or survey mode, the primary metric of interest is survey speed, which favors an LNSD (large number of small diameter dishes) design.

\section{Complementarity} 

The ability to access baseband data is an important capability not just for SETI, but for fast transient applications such as pulsar and FRB studies. Baseband products can also be used for spectral line or other studies, even regenerating higher-level data products to custom specifications from archival data long after the observations are taken (assuming these data are stored).

The proposed search will also provide a superb dynamic catalog of RFI, classified using machine learning algorithms, publicly accessible to other science users. To avoid inadvertent flagging of signals of interest to SETI, the ability to bypass any automatic RFI excision built in to the system is important.

By 2025, Breakthrough Listen \citep{worden:17} will have completed an unprecedented 10-year SETI survey. Even if Listen, or other surveys such as those that are ongoing at the Allen Telescope Array \citep{harp:16}, do not confirm the existence of technological civilizations beyond Earth, they will constrain the parameter space in which such civilizations may exist, motivating a push to higher sensitivities and different frequencies. In the event of an ETI detection, or of the detection of biosignatures or other contingent results from JWST \citep{greene:16}, TESS \citep{tess}, or other planet finding and characterization experiments, ngVLA will provide critical capabilities for follow-up observations.


\acknowledgements Funding for Breakthrough Listen research is provided by the Breakthrough Prize Foundation. 

\bibliography{seti}  



\end{document}